\documentclass{article} 
\usepackage{_ICLR_package/iclr2027_conference,times}

\usepackage{amsmath,amsfonts,bm}

\def\eqref#1{equation~\ref{#1}}

\def\1{\bm{1}}

\DeclareMathAlphabet{\mathsfit}{\encodingdefault}{\sfdefault}{m}{sl}
\SetMathAlphabet{\mathsfit}{bold}{\encodingdefault}{\sfdefault}{bx}{n}

\usepackage{hyperref}
\usepackage{url}
\usepackage{xspace}
\usepackage{wrapfig}
\usepackage{tabularx}
\usepackage{array}
\usepackage{algpseudocode}
\usepackage{algorithm}

\usepackage{booktabs}
\usepackage{multirow}
\usepackage{amsfonts}
\usepackage{graphicx}  

\usepackage{amssymb}
\usepackage{colortbl}
\usepackage{pifont}
\usepackage{makecell}
\usepackage{enumitem} 

\makeatletter
\@ifpackageloaded{xcolor}{}{\usepackage{xcolor}}
\@ifpackageloaded{booktabs}{}{\usepackage{booktabs}}
\@ifpackageloaded{tabularx}{}{\usepackage{tabularx}}
\@ifpackageloaded{array}{}{\usepackage{array}}
\@ifpackageloaded{amsmath}{}{\usepackage{amsmath}}
\@ifpackageloaded{amssymb}{}{\usepackage{amssymb}}
\@ifpackageloaded{listings}{}{\usepackage{listings}}
\@ifpackageloaded{tcolorbox}{}{\usepackage{tcolorbox}}
\@ifpackageloaded{needspace}{}{\usepackage{needspace}}
\@ifpackageloaded{enumitem}{}{\usepackage{enumitem}}
\@ifpackageloaded{hyperref}{}{\usepackage{hyperref}}
\makeatother
\tcbuselibrary{breakable,skins}

\newenvironment{iaCTable}
  {\begingroup\small
   \setlength{\tabcolsep}{4pt}%
   \renewcommand{\arraystretch}{1.08}}
  {\endgroup}

\newtcolorbox{iaCBox}[1]{
  enhanced,breakable,
  colback=black!2,colframe=black!42,boxrule=0.35pt,arc=0pt,
  left=6pt,right=6pt,top=4pt,bottom=4pt,
  before skip=5pt,after skip=5pt,
  title={#1},colbacktitle=white,coltitle=black,
  fonttitle=\small\bfseries,fontupper=\small,
  titlerule=0.35pt,pad at break*=2pt
}

\lstdefinestyle{iaCCode}{
  basicstyle=\ttfamily\fontsize{8.3}{9.7}\selectfont,
  columns=fullflexible,keepspaces=true,breaklines=true,
  breakatwhitespace=false,showstringspaces=false,
  aboveskip=3pt,belowskip=3pt,frame=none,
  xleftmargin=0pt,xrightmargin=0pt,tabsize=2
}

\newenvironment{iaCPrompt}
  {\begin{iaCBox}{Index-Native Corpus Agent Instructions}%
   \fontsize{9}{10.6}\selectfont
   \setlength{\parskip}{3pt}%
   \setlist[itemize]{leftmargin=1.2em,itemsep=1pt,parsep=0pt,topsep=2pt}}
  {\end{iaCBox}}
\newcommand{\iaCPromptHead}[1]{%
  \par\addvspace{4pt}\noindent\textbf{#1}\par\nobreak
}

\makeatletter
\newcommand{\iaContentsSection}[1]{%
  \par\addvspace{8pt}%
  \begingroup
    \hypersetup{hidelinks}%
    \@dottedtocline{1}{0pt}{2.3em}%
      {\hyperref[#1]{\bfseries\numberline{\ref*{#1}}\nameref*{#1}}}%
      {\hyperref[#1]{\bfseries\pageref*{#1}}}%
  \endgroup
}
\newcommand{\iaContentsSubsection}[1]{%
  \par\addvspace{2pt}%
  \begingroup
    \hypersetup{hidelinks}%
    \@dottedtocline{2}{1.6em}{3em}%
      {\hyperref[#1]{\numberline{\ref*{#1}}\nameref*{#1}}}%
      {\hyperref[#1]{\pageref*{#1}}}%
  \endgroup
}
\newenvironment{iaAppendixContents}
  {\begingroup
   \hypersetup{hidelinks}%
   \parindent=0pt\parskip=0pt
   \c@tocdepth=2\relax
   \def\@pnumwidth{2.3em}%
   \def\@tocrmarg{3.3em}%
   \def\@dotsep{4.5}%
  }
  {\par\endgroup}
\makeatother

\ifdefined\DeclareUnicodeCharacter
  \makeatletter
  \@ifpackageloaded{wasysym}{}{\usepackage[nointegrals]{wasysym}}
  \makeatother
  \DeclareUnicodeCharacter{266B}{\twonotes}
\fi

\title{From Delivery to Stateful Exploration:\\
Rethinking the Index for Agentic Search}

\renewcommand{\thefootnote}{\fnsymbol{footnote}}
\author{Deogyong Kim, Sunghwan Kim, Sangam Lee, Wonjae Lee, Dongha Lee\footnotemark[2] \\
Department of Artificial Intelligence\\
Yonsei University\\
Seoul, Republic of Korea \\
\texttt{\{legenduck, happysnail06, salee, dnjswo0926, donalee\}@yonsei.ac.kr} \\
}

\iclrfinalcopy 
\begin{document}
\newcommand{\proposed}{\textsc{IndexAct}\xspace}
\newcommand{\red}[1]{\textcolor{red}{#1}}
\newcommand{\blue}[1]{\textcolor{blue}{#1}}

\newcommand{\sieve}{\textsc{Sieve}\xspace}
\newcommand{\dci}{DCI-Agent-Lite\xspace}
\newcommand{\drdci}{DR-DCI\xspace}
\newcommand{\rarg}{RARG\xspace}
\newcommand{\piserini}{\textsc{Pi-Serini}\xspace}

\maketitle
\footnotetext[2]{Corresponding author.}
\renewcommand{\thefootnote}{\arabic{footnote}}
\begin{abstract}
Recent advances in agentic search have given large language model (LLM) agents finer control over corpus exploration.
However, search interfaces often return matching passages even when feedback about the candidate set would suffice for the next decision, coupling candidate refinement with source-text exposure.
We propose \proposed, an interface for Index-Native Corpus Interaction that separates candidate-set refinement from text inspection.
Agents construct and manipulate persistent candidate sets through lexical conditions and set operations over an inverted index, receiving reusable state references and statistics such as candidate counts rather than matching passages.
This feedback guides further refinement, while separately requested passages provide new clues or evidence that can inform subsequent operations on retained candidate sets.
Experiments on five benchmarks spanning agentic search and multi-hop question answering show that \proposed outperforms the evaluated baselines on each benchmark.
On BrowseComp-Plus, it also achieves higher evidence coverage with a smaller average live context than terminal-based corpus interfaces, and maintains answer accuracy as the corpus expands.
Further analyses suggest that informative refinement feedback and state reuse support continued evidence discovery, while shorter contexts or fewer search steps alone do not ensure better performance.
\href{https://github.com/legenduck/IndexAct}{[CODE]}.
\end{abstract}

\section{Introduction}
\label{sec:introduction}
Search enables large language model (LLM) agents to explore external corpora by submitting queries, inspecting results, and deciding what to do next \citep{yao2023react,jin2025search}.
Conventional retrieval systems make this process efficient by using an index to retrieve and rank candidate documents \citep{lewis2020retrieval,karpukhin2020dense} and returning selected results as text.
An agent can use these results to gather evidence, discover new terms, and revise its queries \citep{trivedi2023interleaving,shao2023enhancing}.
The search interface therefore shapes not only which documents the agent encounters, but also what information enters its context as it decides how to continue.

Recent work has expanded agents’ control over how searches are constructed and revised.
Direct Corpus Interaction (DCI) \citep{li2026beyond} gives agents direct access to raw corpora and general-purpose tools, demonstrating the value of applying exact constraints and inspecting local context during search and verification.
Such control makes observations more important, as each search outcome informs the agent’s next action.
This raises a complementary question: what feedback helps agents decide their next action, and when does that decision require reading document text?

During exploration, an agent may need to assess the effect of a search condition without yet reading the matching content.
For example, when searching for reports about an event, the agent may first want to know how much adding a location narrows the candidate set, or whether it leaves no matches.
Statistical information such as candidate counts \citep{tanin2000facilitating} can inform whether to retain or revise that condition \citep{yang2026superintelligent} without requiring the agent to inspect the reports.
When a search interface returns matching text at each step, however, the agent receives content even when feedback about the candidate set would suffice for the immediate decision.
As such passages accumulate, they occupy the agent’s context even when the current search decision depends primarily on information about the candidate set.
Extending the index beyond result selection to expose candidate-set feedback without automatically returning matching text would let agents refine candidate sets and request relevant passages when needed to discover new clues or verify evidence.

We introduce \proposed, an interface for Index-Native Corpus Interaction that separates candidate-set refinement from text inspection.
Agents construct and refine candidate sets by applying lexical conditions and set operations over the index.
Each new set is returned as a reusable reference with statistics such as its size, rather than matching text.
With the returned reference and statistics, the agent can refine the set further, combine it with a retained set, or request relevant passages.
Retaining intermediate sets also allows the agent to revisit an earlier search state and explore another direction.

\proposed keeps intermediate candidate sets available for further exploration without automatically exposing their contents, letting agents choose what information to inspect at each step.
Agents can use index feedback to assess the effect of a search condition and inspect local context when they need new clues or supporting evidence.
Reading and refinement can alternate throughout exploration: information learned from a passage can guide subsequent operations on retained candidate sets.
By connecting reusable search states, index-based feedback, and agent-directed reading, \proposed gives agents control over both how their search develops and when source text enters their context.

We evaluate \proposed on five benchmarks spanning agentic search and multi-hop question answering.
\proposed achieves the highest task performance on all five, and our agentic-search evaluation shows that it achieves higher evidence coverage with less live context than terminal-based corpus interfaces.
Our analyses suggest that effective context use depends on supporting continued exploration, rather than simply minimizing context length or shortening search trajectories.
Making the effects of search conditions on candidate sets visible and keeping intermediate candidates available for reuse help agents revise their searches, combine earlier results, and seek further evidence.
This support allows candidate exploration to continue without automatically introducing matching passages at each refinement step.
Together, these findings demonstrate that index-native corpus interaction can support effective evidence discovery while keeping source-text exposure selective.

The main contributions of our work are summarized as follows:
\begin{itemize}[leftmargin=1.5em, labelsep=0.5em, itemsep=1pt]
\item
We analyze what enters agent context and when source text enters during evidence discovery, motivating interfaces that support candidate exploration without automatic source-text delivery.
\item
We propose \proposed, an interface for Index-Native Corpus Interaction that lets agents construct, refine, and reuse candidate states through index-based operations, observe their effects through statistical feedback, and inspect source text when needed.
\item
We demonstrate the effectiveness of \proposed across agentic search and multi-hop question answering, and show that informative refinement feedback and reusable candidate states support continued evidence discovery while keeping source-text exposure selective.
\end{itemize}

\section{Related Work}
\label{sec:relatedwork}
\paragraph{Agentic and Long-Horizon Search.}
Agentic search extends retrieval into multi-turn interaction by interleaving reasoning and retrieval \citep{yao2023react,trivedi2023interleaving}. 
Recent work develops learned multi-turn search policies \citep{song2025r1,jin2025search,sun2025zerosearch} and longer investigations that combine repeated search, evidence inspection, and synthesis \citep{li2026webthinker}. 
Retrieved information therefore serves as evidence for answering and as an observation guiding subsequent actions. 
As search trajectories lengthen, context-management methods reorganize working memory or prune stale observations to manage accumulating observations and intermediate reasoning \citep{lu2026longseeker,zhang2026masking}. 
Complementing work on managing accumulated interaction histories, we study what information corpus-access operations return to agents as observations.

\paragraph{Retrieval Interfaces for Agentic Search.}
Corpus interfaces determine the form and granularity of information entering agent context. Recent interfaces give agents finer control over retrieval and evidence access: Interact-RAG~\citep{hui2026interact} exposes retrieval strategies and scope, 
PI-SERINI~\citep{hsu2026rethinking} separates cached-ranking browsing from document reading, and \sieve~\citep{wang2026search} combines Boolean candidate selection, result-card inspection, and selective section fetching. 
\piserini and \sieve therefore avoid automatically exposing full documents, using excerpts or result cards to support decisions about whether and where additional source text should be read. 
\proposed extends this separation to the feedback returned by candidate-set refinement, allowing agents to assess refinement results and operate on retained candidate sets without automatically receiving matching passages.

\paragraph{Direct Corpus Interaction.}
DCI exposes raw corpora through a general-purpose terminal instead of a retriever, enabling composable exploration and localized evidence inspection \citep{li2026beyond}.
However, full-corpus interaction becomes increasingly costly and noisy as corpora grow. 
RISE and \drdci use retrieval to maintain bounded document workspaces for local interaction \citep{zhuang2026towards,lu2026dr}, while \rarg~\citep{li2026new} carries relevance into corpus interaction by prioritizing traversal and matched evidence; these approaches keep retrieved candidates available outside the model context but conduct subsequent fine-grained exploration over materialized documents or file scopes. 
\proposed retains fine-grained evidence access while making candidate sets persistent operands for index-based refinement and composition: these operations return reusable references and statistics rather than source text, which agents inspect separately when needed.

\section{Preliminary Analysis: Context Use During Evidence Discovery}
\label{sec:preliminary_analysis}
Before introducing \proposed, we first examine how existing search agents
use their context while searching for evidence.
Our analysis addresses two questions:

\begin{itemize}[leftmargin=1.5em,topsep=0pt,partopsep=0pt,itemsep=3pt,parsep=0pt]
\vspace{-0.1cm}
    \item \textbf{Question I:}
    What constitutes the context introduced during agentic search?
    \item \textbf{Question II:}
    When is non-evidence source text introduced during evidence discovery?
\end{itemize}
\vspace{-0.3cm}
\paragraph{Agents and Evaluation Setting.}
We analyze context use under two representative corpus-access paradigms on BrowseComp-Plus (BC+)~\citep{chen2025browsecomp}, a fixed-corpus benchmark for iterative search and evidence gathering.
Retrieval Agent (BM25), a ReAct-style agent \citep{yao2023react} based on the official BC+
implementation, uses ranked BM25 results.
\dci~\citep{li2026beyond} replaces the retriever with general-purpose commands such as \texttt{grep} and \texttt{read} for fine-grained corpus access.
We evaluate both agents on the same fixed 100-question subset using GPT-5.4-nano \citep{openai2026gpt54} with high reasoning effort.
Measurement details appear in Appendix~\ref{app:preliminary_analysis_details}.

\subsection{Analysis I: What Constitutes the Agent's Context?}
\label{sec:preliminary_context_composition}

\paragraph{Setup.}
We analyze text retained in each trajectory's final logged model request, excluding the original question, system instructions, and tool specifications.
We group source text by provenance: gold documents, other annotated evidence documents, and documents outside the annotated evidence set.
\textit{Other} includes all remaining interaction text, such as agent-generated messages and non-source content returned by tools.
For each agent, we report the average token composition across trajectories.


\begin{wrapfigure}{r}{0.52\textwidth}
    \centering
    \vspace{-1.2em}
    \includegraphics[width=\linewidth]
    {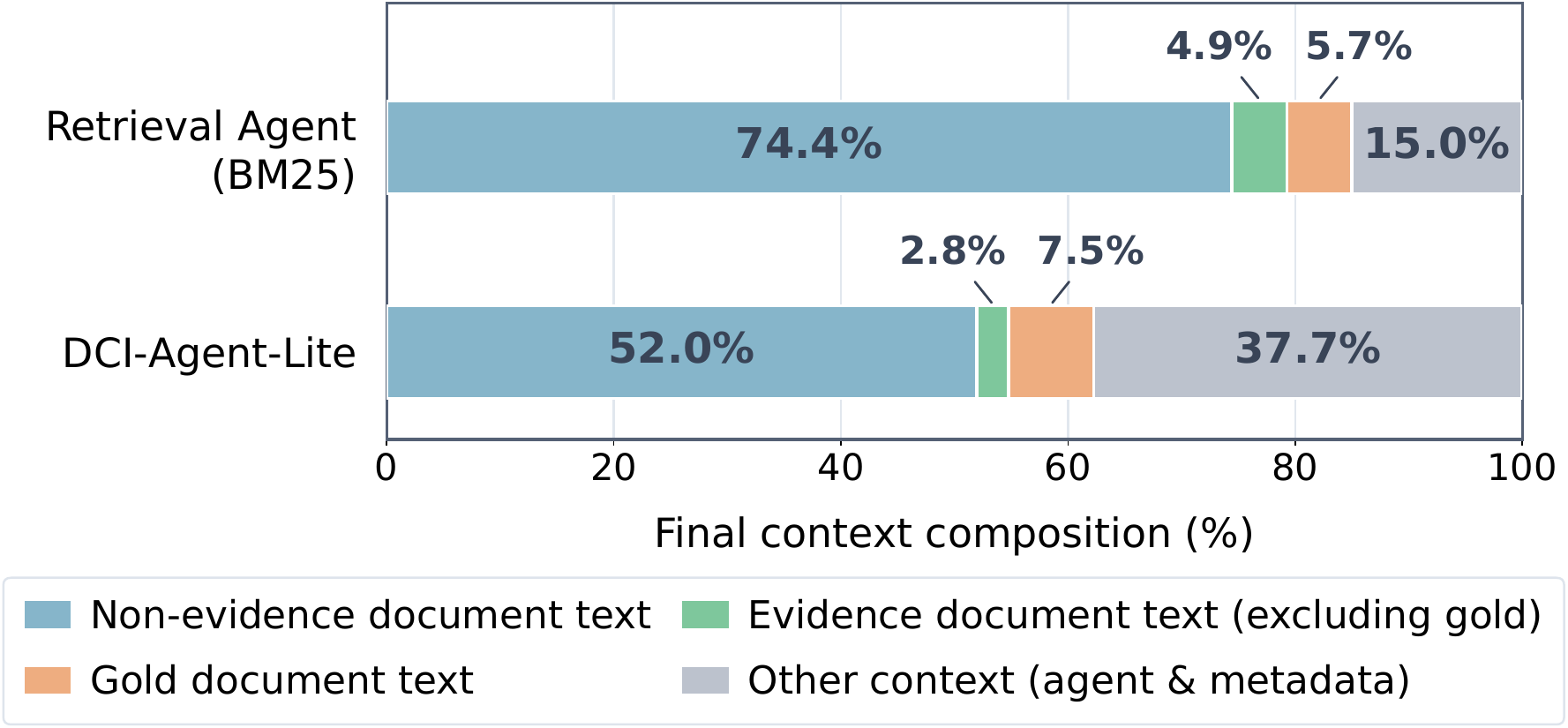}
    \vspace{-0.6cm}
    \caption{
    \textbf{Final context composition.}
    Mean per-trajectory token shares for both agents.
    }
    \label{fig:preliminary_context_composition}
    \vspace{-0.4cm}
\end{wrapfigure}

\paragraph{Corpus observations constitute a substantial share of the retained context.}
Figure~\ref{fig:preliminary_context_composition} shows the final retained context composition for Retrieval Agent (BM25) and \dci.
Source-text observations account for 85.0\% and 62.3\% of the retained context, respectively.
Notably, text from sources outside the annotated evidence set alone accounts for 74.4\% of the context in Retrieval Agent (BM25) and 52.0\% in \dci.
The composition alone does not establish that these observations are unnecessary: text from unannotated sources may reveal new entities, help reject candidates, or provide clues for subsequent searches.
We next examine when this text is observed during evidence discovery.

\subsection{Analysis II: When Is Non-Evidence Text Introduced?}
\label{sec:preliminary_evidence_exposure}

\paragraph{Setup.}
Using the same runs, we focus on trajectories that eventually reach an annotated evidence source.
For each trajectory, we identify the first model request that includes source text from an annotated evidence document.
We then compare the amount of deduplicated non-evidence source text observed up to and including this request with that observed in subsequent requests.
For each trajectory, we compute the corresponding shares and report their average for each agent.

\paragraph{Non-evidence exposure around evidence discovery.}\begin{wrapfigure}{r}{0.52\textwidth}
    \centering
    \includegraphics[width=\linewidth]
    {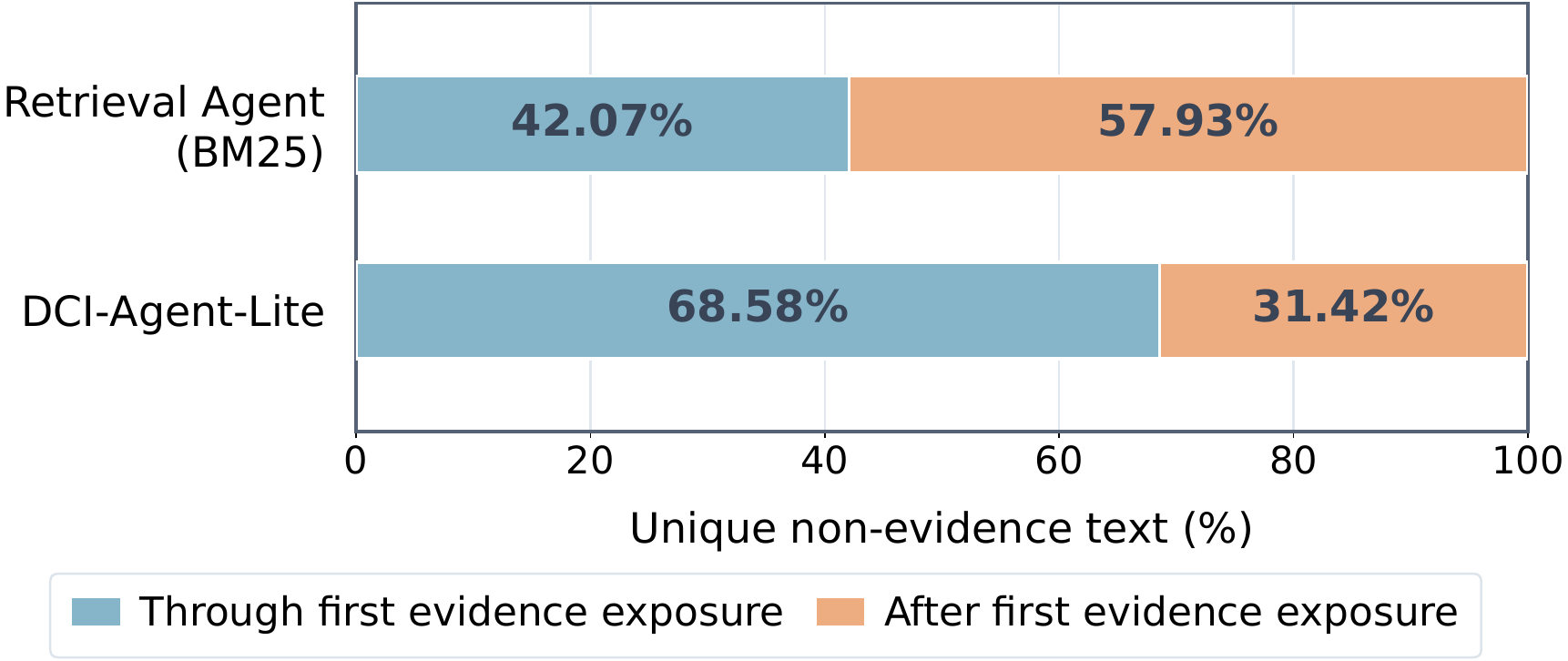}
    \vspace{-0.6cm}
\caption{
    \textbf{When non-evidence source text is observed.}
    Mean per-trajectory token shares.
}
    \label{fig:preliminary_evidence_exposure}
    \vspace{-0.4cm}
\end{wrapfigure}
\dci and Retrieval Agent (BM25) observe, on average, 68.58\% and 42.07\% of their non-evidence source text, respectively, up to and including the first model request containing source text from an annotated evidence document (Figure~\ref{fig:preliminary_evidence_exposure}).
This contrast may reflect how the two access paradigms support continued search.
Retrieval Agent (BM25) can use discovered clues to refine subsequent queries, but each search still exposes retriever-selected text, which may include non-evidence sources.
\dci can instead use discovered clues to guide targeted searches and local inspection.
Its smaller post-evidence share of non-evidence text is consistent with this fine-grained mode of corpus interaction.
The same pattern also points to initial evidence discovery as an important stage for managing source-text exposure, even with fine-grained corpus access.
\vspace{-0.1cm}
\subsection{Insights from Preliminary Analysis}
\label{sec:preliminary_insights}

Our analyses examine \textit{what constitutes the agent's context} and \textit{when source text is observed during search}.
In particular, a substantial share of non-evidence source text is observed while searching for the first evidence source.
Corpus-access design should therefore consider not only how to provide information effectively, but also how to support the search for evidence.

Rather than reducing exploration or restricting necessary reading, we focus on using context effectively to support exploration and the use of evidence.
To this end, we separate decisions about candidate exploration and refinement from decisions about introducing source text into context.
Candidates remain available in the index for exploration, while source text is read selectively to discover new clues or verify evidence.
We aim to reduce the source-text exposure incurred when testing search hypotheses, leaving more context available for necessary reading and further exploration.

\section{\proposed}
\label{sec:design}
\begin{figure*}[t]
    \centering
    \includegraphics[width=0.95\linewidth]{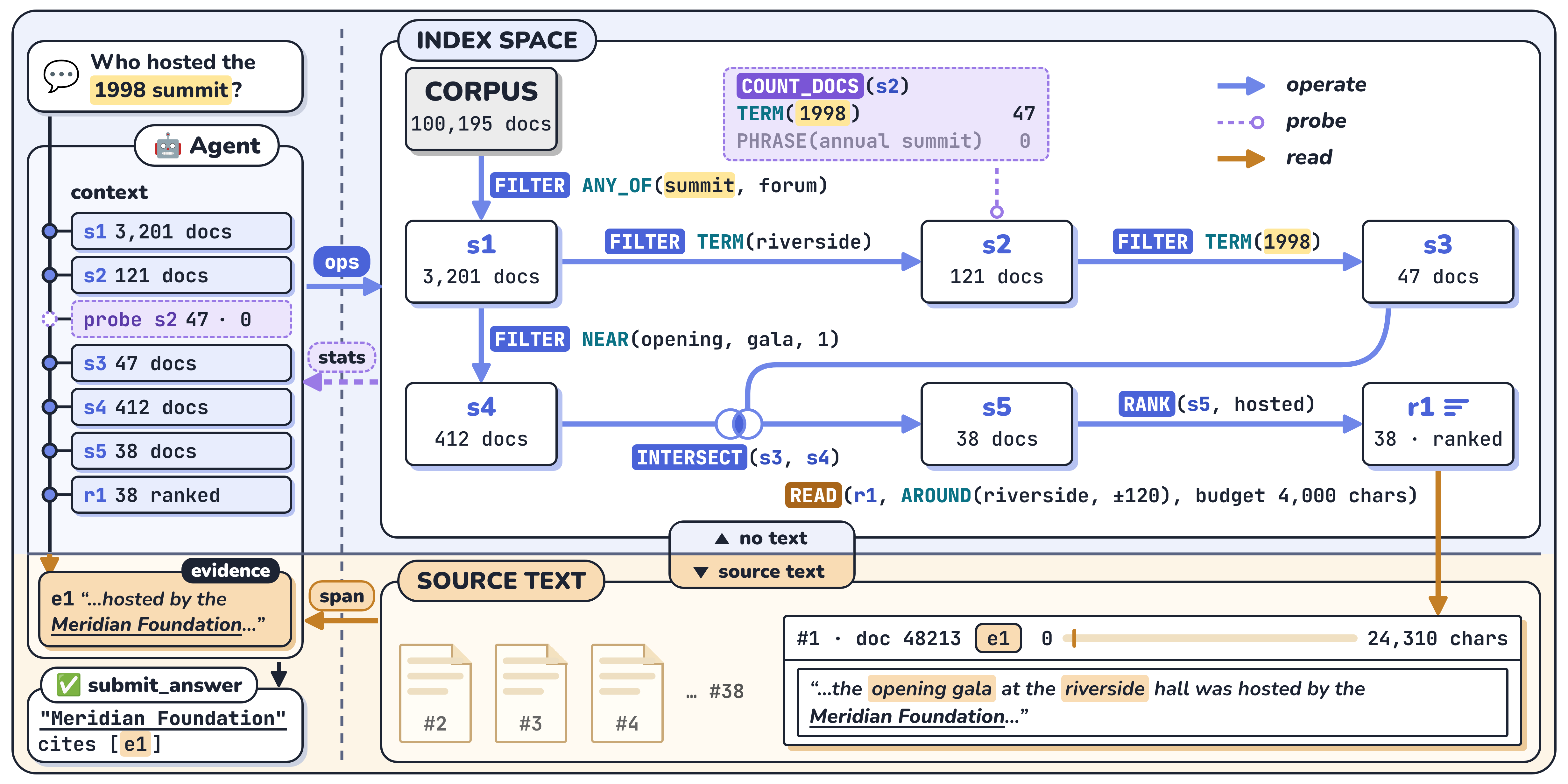}
    \vspace{-0.3cm} 
    \caption{Overview of \proposed. Candidate exploration is performed over reusable states in the index, where operations return state references and statistical feedback without source text. The agent explicitly reads selected source-text regions when it needs new clues or evidence verification.}
    \label{fig:main_fig}
\end{figure*}

We introduce \proposed, an interface for Index-Native Corpus Interaction that provides agents with reusable candidate states and operations to manipulate and inspect them. Agents construct candidates through index-based search and set operations, and reuse the results in subsequent exploration. They can prioritize candidates, inspect statistical feedback, and read selected source-text regions when needed. The operation grammar and detailed execution rules are provided in Appendix~\ref{app:interface_spec}.

\subsection{Overview of Index-Based Agent Interaction}
\label{sec:candidate_states}

\proposed groups candidate documents into a state that can be manipulated throughout subsequent exploration. Its basic form, \texttt{SetRef}, is an unordered set state that records which documents are candidates. The initial candidate state \texttt{CORPUS}, containing the entire corpus, serves as the starting point for exploration. State-producing operations create new states rather than overwriting existing ones, and the agent uses these states as inputs to subsequent operations. It can therefore apply different conditions to earlier candidates or recombine results obtained through different searches.

As shown in Figure~\ref{fig:main_fig}, the agent selects a candidate state and its next action based on the question and previous observations. Candidate-state operations construct candidates by evaluating search conditions through the inverted index, combine candidate sets, and determine inspection order (Section~\ref{sec:search_operations}). State-relative statistics inform subsequent search decisions through candidate-set size and the effects of conditions (Section~\ref{sec:statistical_feedback}). Fine-grained source access lets the agent read around matching locations or within specified regions when needed to examine clues in context (Section~\ref{sec:evidence_access}).

Throughout this process, candidate states remain in the execution environment. The agent can reuse the same state as input to subsequent exploration while selectively inspecting information about its candidates through statistics or source text. Continued use of a candidate state therefore does not require reading all of its documents. Multiple operations determined in advance can be executed together through \texttt{EXECUTE}, and the agent selects its next action based on the returned observations.

\subsection{Candidate-State Operations}
\label{sec:search_operations}

Candidate-state operations address three decisions: which documents to retain, how to combine results, and which documents to inspect first. The operations draw on Boolean retrieval and set algebra for filtering and set composition~\citep{Codd1972RelationalCO}, positional indexing for phrase and proximity search~\citep{Manning_Raghavan_Schütze_2008}, and term-based ranking~\citep{Robertson2009ThePR}. The design aims to let agents draw on prior knowledge of search and data manipulation to construct searches.

\paragraph{Condition filtering.}
\texttt{FILTER} creates a new state containing documents that satisfy a condition within an agent-selected set.
The input condition combines the following expressions. Boolean operators (\texttt{AND}, \texttt{OR}, and \texttt{NOT}) combine required clues, alternatives, and exclusions. 
\texttt{TERM} specifies term occurrence. Positional conditions constrain the order and spacing of clues within a document, using \texttt{PHRASE} for ordered adjacency and \texttt{NEAR} for proximity. \texttt{ANY\_OF} matches locations where any of its alternatives, such as a term or phrase, occurs.
Boolean and positional conditions distinguish requiring clues to occur in the same document from requiring them to occur near each other.

Let $h$ be the input state reference, $S_h$ its document set, $p$ a condition, and $\mathcal{I}$ the inverted index. Then
\begin{equation}
S_{h'}=F_{\mathcal{I}}(S_h,p)
=\{d\in S_h\mid d\models_{\mathcal{I}}p\},
\label{eq:filter}
\end{equation}
where $h'$ references the resulting state and $d\models_{\mathcal{I}}p$ means that document $d$ satisfies condition $p$.

\paragraph{Set composition.}
The agent combines set states through intersection (\texttt{INTERSECT}), union (\texttt{UNION}), and difference (\texttt{DIFFERENCE}).
Results for different clues can be retained separately, inspected independently, and recombined as needed. For example, the agent can retain candidates shared by multiple clues, merge candidates found through aliases, or exclude identified distractors.

\paragraph{Ranking.}
\texttt{RANK} determines the inspection order without removing candidates, producing a ranked state, \texttt{RankedRef}, containing the same documents. The agent specifies ranking terms (\texttt{TERM}) and weights. For each document in the candidate set, the executor computes each term's BM25 contribution and sums these contributions using the agent-specified weights to determine the ranking.

This separation allows the agent to distinguish requirements from preferences. Entity names or phrase and proximity relations that the agent considers mandatory can be used as filter conditions, while auxiliary terms that suggest relevance but are unsuitable as requirements can guide ranking. Documents lacking those auxiliary terms remain candidates; only their inspection order changes.

\paragraph{Ranked-state operations.}
For a ranked state, \texttt{TOPK} selects up to the top $k$ candidates, while \texttt{RESTRICT} retains documents also in a specified set, preserving their order.
For subsequent filtering or set composition, \texttt{AS\_SET} produces a new unordered set state containing the same documents.

\subsection{State-Relative Statistical Feedback}
\label{sec:statistical_feedback}

Statistical feedback helps assess the effects of search conditions without reading candidate source text, informing whether to revise a condition, broaden the scope, or inspect source evidence. Result-size feedback has long supported exploratory query formulation~\citep{tanin2000facilitating}, and recent agentic retrieval uses corpus statistics to validate proposed search vocabulary~\citep{yang2026superintelligent}. The current feedback consists of candidate-set sizes and counts of documents matching each condition. Statistical observations do not automatically include document lists or source-text passages.

\paragraph{State statistics.}
Creating a state also returns its candidate count. By comparing counts before and after an operation, the agent can determine whether the result is empty, how much a condition reduces the candidates, or how much combining results expands the set of candidate documents.

\paragraph{Condition probing.}
\texttt{COUNT\_DOCS} evaluates a condition within a retained candidate state and returns the number of matching documents without registering a new candidate state. Comparing match counts for individual clues and their combinations helps the agent identify and adjust restrictive conditions. Testing the same condition on the current candidates and on \texttt{CORPUS} reveals whether matches exist beyond the current scope, helping it decide whether to broaden the search.
The agent can then choose which results to retain as candidate states for subsequent exploration.

\subsection{Fine-Grained Evidence Access}
\label{sec:evidence_access}

Fine-grained source access lets the agent examine how the clues are used in their original passages and assess whether the context surrounding a lexical match supports the intended interpretation.

\paragraph{Source-text reading.}
The agent calls \texttt{READ} on a candidate state or an individual document and selects text using its region argument.
For ranked states, reading follows their established order.

To read around matches, the agent uses \texttt{AROUND} as the region argument and specifies an anchor expression and the context extent before and after each selected match. The executor uses the index to locate matches and return surrounding source text. For example, the agent can reuse a \texttt{NEAR} expression from candidate search as the anchor to examine what relationship the nearby clues express. Changing the anchor allows the agent to investigate other clues within the same candidate state.

The agent uses \texttt{DOCUMENT} as the region argument to read a full document, or \texttt{RANGE} to read a specified region in an individual document.
Read results identify the source document and passage location, so the agent can inspect adjacent context or revisit the passage without locating it again.

\section{Results}
\label{sec:results}
\subsection{Experimental Setup}
\label{sec:experimental_setup}

\paragraph{Benchmarks.}
We evaluate \proposed on all 830 questions in BrowseComp-Plus (BC+)~\citep{chen2025browsecomp}, our primary benchmark for agentic search.
We additionally evaluate multi-hop QA on HotpotQA~\citep{yang2018hotpotqa}, the answerable version of MuSiQue~\citep{trivedi2022musique}, 2WikiMultiHopQA (2Wiki)~\citep{ho2020constructing},
and Bamboogle~\citep{press2023measuring}.
Following the evaluation scale used in prior multi-hop QA studies~\citep{trivedi2023interleaving,jeong2024adaptive}, we use fixed 500-question subsets of HotpotQA, MuSiQue, and 2Wiki, and evaluate all 125 Bamboogle questions.
The same question subsets are used across all methods.
BC+ uses its released corpus.
For the four QA benchmarks, all methods use the full Wikipedia-18 corpus~\citep{karpukhin2020dense},
containing approximately 20M passages, as the underlying corpus.
More details are provided in Appendix~\ref{app:datasets}.

\paragraph{Baselines.}
To evaluate the index-native candidate manipulation and selective reading enabled by \proposed, we compare against methods spanning fine-grained corpus interaction and selective evidence access.
We include \dci~\citep{li2026beyond}, \drdci~\citep{lu2026dr}, and
\rarg~\citep{li2026new} as terminal-based alternatives for corpus exploration.
We also include \piserini~\citep{hsu2026rethinking} and \sieve~\citep{wang2026search} to compare against interfaces that already separate retrieval from selective evidence access.
As conventional agentic-search baselines, we include Retrieval Agent (BM25) and Retrieval Agent (Dense), which use the same ReAct-style search--read loop~\citep{yao2023react}
and differ only in the retrieval backend.
More detailed descriptions of each baseline are provided in Appendix~\ref{app:baselines}.

\paragraph{Implementation details.}
We use GPT-5.4 nano with high reasoning effort across the compared interfaces.
We run \proposed in a lightweight Pi-based agent harness~\citep{zechner2026pi} and expose its operations through structured tool calls.
The backend is implemented as a Java service over Apache Lucene, with positional postings for lexical and proximity operations.
Candidate-set states are maintained server-side and exposed to the agent through reusable references.
We impose a 300-turn limit per question and a 30-second timeout per tool call.
Baseline-specific settings, including retrievers and indexes, are described in Appendix~\ref{app:baselines}, and the full operation specification and agent instruction are provided in Appendices~\ref{app:interface_spec} and ~\ref{app:prompts}.

\paragraph{Metrics.}
We report Accuracy on all benchmarks. For each question, an LLM judge determines whether the final answer from the agent's response is semantically equivalent to the reference answer.
On BC+, we additionally report Average Live Context and Evidence Coverage.
Average Live Context is the per-query mean number of dynamic-history tokens supplied to decision-making model calls after truncation or compaction, excluding fixed prompts, tool schemas, and the question.
Evidence Coverage is the fraction of annotated evidence documents whose source text enters an agent input during the trajectory; candidate membership alone does not count as observation.
Query-level metrics are averaged over questions, including failed and budget-exhausted runs.
Answer-scoring procedures, the judge model and prompt, and metric definitions are provided in Appendix~\ref{app:evaluation_metrics}.

\subsection{Main Results}
\label{sec:main_results}


\begin{table}[t]
\centering
\small
\setlength{\tabcolsep}{6pt}   
\renewcommand{\arraystretch}{0.9}

\caption{
End-to-end comparison of search agents on BrowseComp-Plus and multi-hop QA benchmarks.
Bold and underlined entries mark the best and second-best in each column, respectively.
}
\label{tab:main_results}
\vspace{0.2cm}

\resizebox{\linewidth}{!}{%
\begin{tabular}{lccccccc}
\toprule

& \multicolumn{3}{c}{\textbf{BrowseComp-Plus}}
& \textbf{MuSiQue}
& \textbf{Hotpot}
& \textbf{2Wiki}
& \textbf{Bamb.} \\

\cmidrule(lr){2-4}
\cmidrule(lr){5-5}
\cmidrule(lr){6-6}
\cmidrule(lr){7-7}
\cmidrule(l){8-8}

\textbf{Method}
& Acc.\,$\uparrow$
& Evi. Cov.\,$\uparrow$
& Avg. Ctx.\,$\downarrow$
& Acc.\,$\uparrow$
& Acc.\,$\uparrow$
& Acc.\,$\uparrow$
& Acc.\,$\uparrow$ \\

\midrule

\multicolumn{8}{l}{\textit{Standard Search Agents}} \\
\addlinespace[2pt]

Retrieval Agent (BM25)
& 30.5 & 36.5 & 31,243
& 37.8 & 68.4 & 63.4 & 68.8 \\

Retrieval Agent (Dense)
& 48.2 & 54.4 & 24,543
& \underline{44.4}
& 69.2
& \underline{69.0}
& \underline{70.4} \\

\midrule

\multicolumn{8}{l}{\textit{Agentic Corpus Interfaces}} \\
\addlinespace[2pt]

\piserini
& 62.7 & 55.4 & \textbf{10,835}
& 41.0 & 68.4 & 67.2 & 68.0 \\

\sieve
& 53.0 & 40.8 & 15,931
& 40.2 & 68.6 & 59.2 & 68.0 \\

\dci
& 63.0 & 43.6 & 30,161
& 42.0 & 68.6 & 63.6 & 63.2 \\

\drdci
& 67.2 & 46.6 & 22,435
& 42.6 & \underline{70.4} & 66.8 & 64.8 \\

\rarg
& \underline{69.4}
& \underline{57.2}
& 27,896
& 43.4 & 68.2 & 67.0 & 64.0 \\

\midrule

\rowcolor{gray!8}
\textbf{\proposed{}}
& \textbf{73.5}
& \textbf{60.7}
& \underline{13,960}
& \textbf{46.2}
& \textbf{71.8}
& \textbf{69.2}
& \textbf{71.2} \\

\bottomrule
\end{tabular}
}
\end{table}

\paragraph{Performance on BC+.}

Table~\ref{tab:main_results} shows that \proposed achieves the highest answer accuracy and Evidence Coverage on BC+.
Among the corpus interfaces that outperform conventional Retrieval Agents, \rarg is the strongest baseline, while \proposed further improves both metrics.
\rarg combines retriever-based prioritization with 
fine-grained source-text exploration by traversing documents in the order induced by the retriever.
However, this also means that which search matches reach the agent first depends on the retriever's ranking.
Because only a bounded number of matches can be returned to the agent, matches from higher-ranked documents can occupy the available output while useful clues in lower-ranked documents remain unobserved~\citep{li2026new}.
In contrast, \proposed lets the agent directly refine and combine retained candidate sets according to its current search conditions, assess their effects through statistical feedback, and separately decide what source text to inspect.
The improvements in both accuracy and Evidence Coverage suggest that, beyond relevance-guided document traversal, explicit control over candidate sets and feedback on their
refinement can further support effective evidence search.

\paragraph{Context use.}
Compared with the terminal-based alternatives, \proposed achieves higher answer accuracy and Evidence Coverage with a lower Average Live Context, about half that of \rarg.
In terminal-based interfaces, each search returns matching text, and these outputs accumulate in the interaction history as exploration proceeds.
\proposed instead keeps the progress of candidate refinement in index-side states and returns references and statistics, so source text enters the context only through explicit reads.
Against the selective-reading baselines, \proposed outperforms \sieve on all three measures, and although \piserini uses even less context, its answer accuracy and Evidence Coverage are lower.
These comparisons highlight the importance of using context effectively for evidence discovery, rather than simply minimizing Average Live Context.

\paragraph{Multi-hop QA.}
The benefits also extend to multi-hop QA over Wikipedia-18.
Here, Retrieval Agent (Dense) remains a strong alternative: it is the highest-scoring baseline on MuSiQue, 2Wiki, and Bamboogle, whereas DR-DCI performs best on HotpotQA.
Strong BC+ performance therefore does not uniformly translate into an advantage on these QA benchmarks.
Nevertheless, \proposed achieves the highest accuracy on all four benchmarks.
Although the margins are smaller than on BC+, these improvements show that the benefits of index-native corpus interaction are not confined to the primary evaluation and extend to multi-hop QA over a different underlying corpus.

\section{Analysis}
\label{sec:analysis}
\subsection{Candidate Refinement and Evidence Acquisition}
\label{sec:observation_analysis}

We examine how refinement feedback affects evidence acquisition and context use, keeping candidate-state operations and explicit reading unchanged.
\textit{Auto Preview} adds bounded source-text snippets to \texttt{FILTER} and set-composition outputs.
\textit{Binary Feedback} replaces exact counts in state-producing outputs and \texttt{COUNT\_DOCS} responses with empty/non-empty indicators.
Alongside accuracy, evidence coverage, and average live context, we report model calls per query and cumulative source tokens returned by tools per query. More details are provided in Appendix~\ref{app:observation_analysis}.

\begin{wraptable}{r}{0.52\textwidth}        
  \centering
  \vspace{-0.76cm}
  \caption{Observation-policy ablation on BC+.}
  \label{tab:observation_ablation}
  \vspace{0.1cm}
  \fontsize{8.5}{9}\selectfont             
  \setlength{\tabcolsep}{2pt}               
  \begin{tabular*}{\linewidth}{@{\extracolsep{\fill}}lccccc@{}}
    \toprule
    Method & Acc.$\uparrow$ & EviCov.$\uparrow$ & AvgCtx$\downarrow$ & Calls & SrcTok.$\downarrow$ \\
    \midrule
    Auto Preview      & 70.0 & 52.5 & 14,842          & 38.5 & 32.0K \\
    Binary Feedback & 62.0 & 55.9 & \textbf{14,492} & 39.8 & 35.2K \\
    \textbf{\proposed{}} & \textbf{74.0} & \textbf{63.9} & 15,107 & 42.9 & \textbf{30.8K} \\
    \bottomrule
  \end{tabular*}
\end{wraptable}

\paragraph{Observation policy.}
Auto Preview uses fewer model calls but lowers both accuracy and evidence coverage (Table~\ref{tab:observation_ablation}).
Average live context remains close to that of \proposed{}, as additional tool-response content is offset by less agent-command history.
The joint reduction in calls and coverage suggests that previews encourage earlier answer commitment before further evidence is sought or verified.
An immediate clue can guide search, but does not establish that the remaining constraints have been satisfied.
\proposed{} instead achieves broader evidence coverage with more model decisions but less source-text exposure. Thus, shorter trajectories do not necessarily yield more effective search.

\paragraph{Statistical feedback.}
Binary Feedback has the smallest average live context, yet exposes the most source text and yields the lowest accuracy.
Its smaller context reflects less accumulated non-source interaction history, not reduced source exposure.
Exact counts reveal whether a condition substantially narrows the candidates or leaves the search largely unchanged, helping the agent decide whether to refine further or read.
The lower accuracy and evidence coverage despite greater source exposure suggest that additional text does not compensate for the loss of this state-level guidance.
Statistical feedback therefore acts as a search-control signal rather than compact metadata.
Together, the comparisons favor informative feedback and selective reading over minimizing working context or interaction count alone.

\subsection{Candidate States and Exploration}
\label{sec:state_analysis}


\paragraph{Candidate states.}\begin{wraptable}{r}{0.59\textwidth}        
  \centering
  \vspace{-2.2em}
  \caption{Candidate-state and operation ablations on BC+.}
  \vspace{0.4em}
  \label{tab:state_ablation}
  \fontsize{8.5}{8.5}\selectfont              
  \renewcommand{\arraystretch}{1}
  \begin{tabularx}{\linewidth}{@{}l@{\hspace{6pt}}*{4}{>{\centering\arraybackslash}X}@{}}
    \toprule
    Method & Acc.$\uparrow$ & EviCov.$\uparrow$ & AvgCtx$\downarrow$ & Calls \\
    \midrule
    \textbf{\proposed{}} & \textbf{74.0} & \textbf{63.9} & 15,107 & 42.9 \\
    \midrule
    w/o State Reuse     & 66.0 & 58.4 & \textbf{14,365} & 42.2 \\
    w/o Set Composition & 62.0 & 53.1 & 16,946          & 54.9 \\
    w/o Positional Ops. & 70.0 & 59.1 & 14,907          & 44.3 \\
    \bottomrule
  \end{tabularx}
  \vspace{-0.5cm}
\end{wraptable}
We examine the contributions of candidate states and index operations by removing each capability separately while retaining the observation policy of \proposed (Table~\ref{tab:state_ablation}).
More details are provided in Appendix~\ref{app:state_analysis}.
Removing state reuse lowers both accuracy and evidence coverage despite a similar number of model calls, suggesting that access to intermediate results supports exploration beyond simply increasing the number of interaction steps.
Removing set composition, while preserving Boolean filtering, causes the largest performance drop and increases both model calls and average live context.
The additional interaction does not recover the loss in evidence coverage, indicating that direct composition provides value beyond expressing multiple conditions
in a new query.
Together, these support treating intermediate candidate sets as reusable and composable search states rather than one-off retrieval outputs.

\paragraph{Positional operations.}
Removing \texttt{PHRASE} and \texttt{NEAR} also lowers accuracy and evidence coverage, despite slightly more model calls and similar average live context.
This suggests that document-level term co-occurrence alone does not provide the same control over evidence search as explicit phrase and proximity conditions.
Positional operations therefore complement Boolean filtering by allowing the agent to specify more precise lexical relationships when selecting candidates.

\subsection{Robustness to Corpus Scale}
\label{sec:corpus_scaling}

\begin{figure*}[t]
    \centering
    \includegraphics[width=0.9\linewidth]{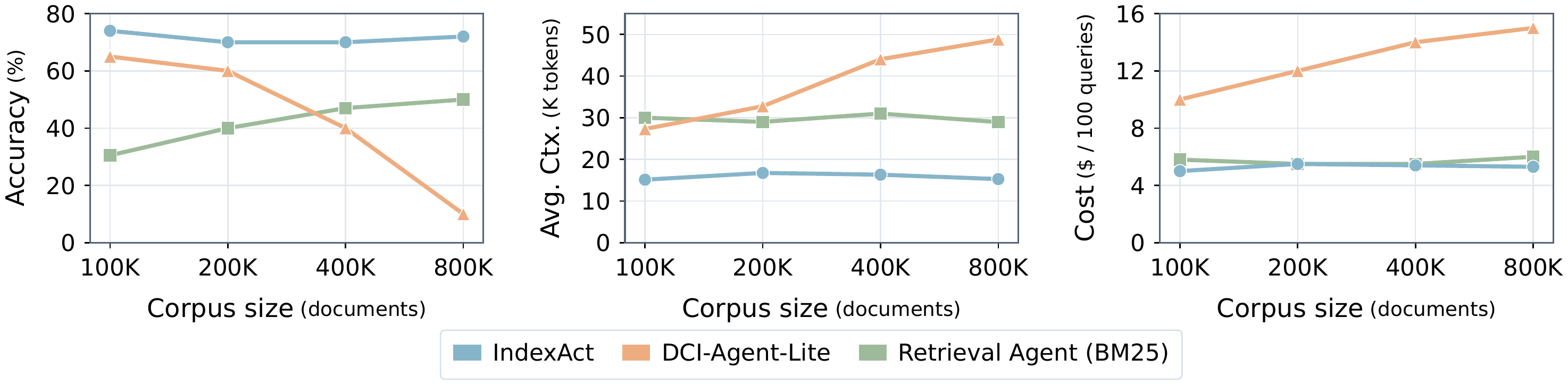}
    \vspace{-0.3cm}
\caption{
Effect of corpus expansion on accuracy, average live context,
and API cost on BC+.
The corpus grows from 100K to 800K documents, with the same
100 questions and evidence.
}
    \label{fig:corpus_scaling}
\end{figure*}

\paragraph{Corpus expansion.}
To examine whether fine-grained corpus exploration can retain the scaling advantages of indexed retrieval, we compare
\proposed with Retrieval Agent (BM25) and DCI-Agent-Lite as the corpus expands.
We keep 100 BC+ questions and their original evidence fixed, expanding the corpus from approximately 100K to 200K, 400K,
and 800K documents using nested samples of FineWeb pages~\citep{penedo2024fineweb}.
Model settings and execution budgets remain fixed within each method across scales.
Corpus construction is detailed in Appendix~\ref{app:corpus_scaling}.

\paragraph{Performance and resource use.}
Figure~\ref{fig:corpus_scaling} shows that \proposed{} maintains
70.0--74.0\% accuracy as the corpus grows eightfold, with little
change in average live context or API cost.
DCI-Agent-Lite instead uses more context and incurs higher costs
while its accuracy declines.
Broad terminal searches may scan more raw text as the corpus grows,
while returned matches and tool-error messages can accumulate
in the interaction history without advancing the search.

\proposed{} evaluates lexical and positional conditions over index postings and retained candidate sets rather than repeatedly
scanning raw documents.
Refinement returns references and statistics, with source text accessed through separate reads.
A larger corpus can therefore increase index-side processing without a corresponding increase in the text returned by each
refinement.
Retrieval Agent (BM25) also exhibits stable resource use, consistent with indexed access, but remains less accurate than
\proposed{} across tested sizes.
These support extending index-based execution to fine-grained candidate exploration while maintaining answer
quality and stable working-context use.

\section{Conclusion}
\label{sec:conclusion}
This paper proposes \proposed, an interface for Index-Native Corpus Interaction that separates candidate-set refinement from source-text inspection.
It lets agents construct, refine, and reuse persistent candidate states through index-based operations, using statistical feedback to guide exploration while selectively reading source text when needed.
Our experiments demonstrate strong performance across agentic search and multi-hop question answering, while further analyses show that informative refinement feedback and reusable candidate states support effective evidence discovery and remain robust as corpus size grows.
We hope that \proposed will contribute to future work on more effective and context-efficient interfaces for agentic search.


\subsection*{AI use statement}
LLMs were used to assist with manuscript writing and editing for clarity and readability, as well as with code implementation for the experiments. 
All AI-assisted text was reviewed by the authors, and all AI-generated code was manually verified before being used in the experiments. 
The authors remain fully responsible for the final content of the manuscript and the code used in this work.

\bibliography{bibliography}
\bibliographystyle{_ICLR_package/iclr2027_conference}


\appendix
\clearpage
\appendix
\begingroup
\hypersetup{hidelinks}
\providecommand{\red}[1]{\textcolor{red}{#1}}
\providecommand{\proposed}{\textsc{IndexAct}}

\pdfbookmark[1]{Contents of Appendix}{indexact.appendix.contents}
\section*{Contents of Appendix}
\begin{iaAppendixContents}
  \hfill\textit{Page}\par

  \iaContentsSection{app:interface_spec}
  \iaContentsSubsection{app:operations_matching}
  \iaContentsSubsection{app:reading_execution}
  \iaContentsSubsection{app:interaction_example}

  \iaContentsSection{app:experimental_setup}
  \iaContentsSubsection{app:datasets}
  \iaContentsSubsection{app:baselines}

  \iaContentsSection{app:evaluation_metrics}
  \iaContentsSubsection{app:scoring_accounting}
  \iaContentsSubsection{app:preliminary_analysis_details}

  \iaContentsSection{app:additional_analyses}
  \iaContentsSubsection{app:observation_analysis}
  \iaContentsSubsection{app:state_analysis}
  
  \iaContentsSection{app:corpus_scaling}
  \iaContentsSection{app:prompts}
\end{iaAppendixContents}
\clearpage

\section{Index-Native Interaction}
\label{app:interface_spec}

The interface separates candidate membership, inspection order, and source-text observation. An unordered \texttt{SetRef} identifies candidate documents, and a \texttt{RankedRef} adds an inspection order. The root \texttt{CORPUS} denotes the entire corpus. State-producing operations create new states without overwriting their inputs and return a reference, count, and state metadata rather than text or membership lists. References remain available within the session until released.

\subsection{Operations and matching semantics}
\label{app:operations_matching}
The grammar uses $S$ for an unordered state, $R$ for a ranked state, and $H$ for either. \texttt{AS\_SET} converts ranked states before unordered-set operations.

\begin{iaCBox}{Candidate-state and expression grammar}
\begin{lstlisting}[style=iaCCode]
State operations and counts
  FILTER(S, p)                         -> SetRef
  INTERSECT(S1, S2) | UNION(S1, S2)     -> SetRef
  DIFFERENCE(S1, S2) | AS_SET(R)        -> SetRef
  RANK(S, score) | TOPK(R, k)           -> RankedRef
  RESTRICT(R, S)                       -> RankedRef
  COUNT(H) | COUNT_DOCS(H, p)           -> integer
Lexical expression
  e ::= TERM(text) | ANY_OF(e1,...,en)
      | PHRASE(e1,...,en)
      | NEAR(e1,...,en, ordered, max_gaps)
Document condition
  p ::= e | AND(p1,...,pn) | OR(p1,...,pn) | NOT(p)
Ranking expression
  score ::= TERM(text) | COMBINE(TERM(t1),...,TERM(tn))
          | WEIGHT(w1,TERM(t1),...,wn,TERM(tn))
\end{lstlisting}
\end{iaCBox}

\paragraph{Filtering, composition, and counts.}
\texttt{FILTER} evaluates a document condition within its input set. In particular, \texttt{NOT} excludes matching documents from that set, not from an implicit corpus-wide search. \texttt{INTERSECT}, \texttt{UNION}, and \texttt{DIFFERENCE} combine set membership without duplicates. \texttt{COUNT} returns the size of an existing state; \texttt{COUNT\_DOCS} returns the number of documents satisfying a condition within that state without registering another state. Neither count operation returns matching passages.

\paragraph{Lexical and positional conditions.}
The analyzer applies Unicode normalization and lowercase mapping without stemming or stopword removal. Each \texttt{TERM} argument must produce exactly one analyzed token, and matching is performed at the token level. \texttt{AND} and \texttt{OR} combine document-level conditions. In contrast, \texttt{ANY\_OF} preserves matching locations from its alternatives, allowing it to occur within a positional expression or serve as a reading anchor. \texttt{PHRASE} requires ordered adjacency. \texttt{NEAR} bounds the number of uncovered token positions between its child matches and optionally requires their submitted order. The same matching semantics apply to candidate selection and \texttt{AROUND} anchors.

\paragraph{Ranking and membership.}
\texttt{RANK} uses a weighted sum of BM25 term contributions, with non-negative weights; \texttt{COMBINE} assigns unit weights. The index implementation uses $k_1=1.2$ and $b=0.75$, with corpus-wide statistics rather than statistics recomputed within each candidate set. \texttt{RANK} changes only the inspection order and preserves all input members. Ties are broken by document-key order. \texttt{TOPK} retains the highest-ranked $k$ members, whereas \texttt{RESTRICT} retains members also found in a supplied set without changing their order. \texttt{AS\_SET} removes ordering but preserves membership.

\subsection{Reading and execution}
\label{app:reading_execution}
\begin{iaCBox}{Source-reading grammar}
\begin{lstlisting}[style=iaCCode]
Source reading
  READ(H, documents, region, budget)
  READ(DocKey, region, budget)
  documents ::= ALL_DOCUMENTS | PAGE(limit, after?)
  region    ::= DOCUMENT | RANGE(start,end)
              | AROUND(e, selector, before, after)
  selector  ::= FIRST | NTH(n) | ALL
  budget    ::= {max_output_codepoints, max_evidence_count}
\end{lstlisting}
\end{iaCBox}

\paragraph{Document and region selection.}
A state read selects all documents or a page without changing candidate membership. Unordered states follow document-key order; ranked states follow their established order. \texttt{DOCUMENT} reads the whole selected document, and \texttt{RANGE} reads a half-open interval in an individually identified document. \texttt{AROUND} locates an anchor match and returns the requested context before and after it. Its selector chooses the first, the $n$th, or all occurrences; overlapping windows are merged when all occurrences are requested. Reading offsets and context extents are measured in raw-text Unicode code points, whereas phrase and proximity conditions use analyzed-token positions.

\paragraph{Returned evidence and output limits.}
Passages include their source document, raw-text offsets, and exact text. To preserve passage integrity, over-budget requests return budget feedback so that the agent can issue a smaller read. A candidate-state reference by itself is not an evidence citation.

\paragraph{Batched operations.}
The \texttt{index\_execute} tool batches non-text operations, and \texttt{index\_read} returns text. An \texttt{EXECUTE} batch supports dependent state operations, while adaptive decisions are made across tool calls. State metadata and lineage can be inspected through \texttt{index\_state} without reading candidate contents.

\subsection{A worked interaction example}
\label{app:interaction_example}

Consider a four-document example corpus: \texttt{doc-a} contains \texttt{a b}, \texttt{doc-b} contains \texttt{a c}, \texttt{doc-c} contains \texttt{c}, and \texttt{doc-empty} is empty. Table~\ref{tab:expanded_example} illustrates the information returned by successive operations. State references ($h$) are shown symbolically, and only the relevant response fields are shown.

\begin{table}[htbp]
\centering
\caption{Candidate feedback and source observation on the four-document example corpus. \texttt{READ} uses the complete-document region and a sufficient output budget.}
\vspace{0.2cm}
\label{tab:expanded_example}
\begin{iaCTable}
\begin{tabularx}{\linewidth}{@{}X l@{}}
\toprule
Operation & Returned information \\
\midrule
\texttt{FILTER(CORPUS, TERM("a"))} & $h_a$, count $2$ \\
\texttt{FILTER(CORPUS, TERM("c"))} & $h_c$, count $2$ \\
\texttt{INTERSECT}$(h_a,h_c)$ & $h_{\cap}$, count $1$ \\
\texttt{DIFFERENCE}$(h_a,h_c)$ & $h_{\setminus}$, count $1$ \\
\texttt{COUNT}$(h_a)$ & $2$; no new state \\
\midrule
\texttt{READ}$(h_{\cap},\texttt{DOCUMENT})$ &
\texttt{doc-b}, $[0,3)$, \texttt{a c} \\
\bottomrule
\end{tabularx}
\vspace{-0.4cm}
\end{iaCTable}
\end{table}

The first five responses contain no source text. Intersection and difference create new candidate states while preserving their input states. Only the final read exposes a passage. The agent could instead inspect the residual branch or apply another condition to an earlier state. A singleton result narrows eligibility but does not by itself establish that the document supports an answer.

\section{Experimental Details}
\label{app:experimental_setup}

\subsection{Datasets and sampling}
\label{app:datasets}

The primary evaluation uses all 830 BC+ questions and the released corpus~\citep{chen2025browsecomp}. For multi-hop QA, we use fixed 500-question subsets of HotpotQA, the answerable version of MuSiQue, and 2WikiMultiHopQA, and all 125 Bamboogle questions~\citep{yang2018hotpotqa,trivedi2022musique,ho2020constructing,press2023measuring}. The 500-question subsets are randomly sampled from the respective datasets. Each subset is held fixed across methods.

All four QA benchmarks use the full Wikipedia-18 corpus, containing approximately 20M passages, rather than task-specific evidence collections~\citep{karpukhin2020dense}. Each comparison uses a common corpus. The preliminary and scaling analyses each use 100 BC+ questions; Appendix~\ref{app:additional_analyses} specifies the ablation subsets.

\subsection{Baselines}
\label{app:baselines}

We compare \proposed with agents that access corpora through ranked retrieval results, direct corpus interaction, or structured search and selective reading. 
All baselines are run with their official implementations, using the same backbone model and execution limits as \proposed. On BC+, sparse and dense retrieval use the BM25 and Qwen3-Embedding-8B~\citep{zhang2025qwen3} indexes released with the benchmark~\citep{chen2025browsecomp}. On Wikipedia-18, we build a BM25 index with the standard Anserini indexing pipeline~\citep{10.1145/3077136.3080721}, using the same BM25 settings as \proposed, and use E5-base-v2~\citep{wang2022text} for dense retrieval, following prior work on this corpus~\citep{jin2025search,li2026beyond}. Baselines that rely on a sparse or dense retriever use these same retrievers on each corpus, so that retriever choice is held fixed across methods.

\paragraph{Retrieval Agent (BM25/Dense).}
Retrieval Agent follows a ReAct-style search--read loop:
the agent submits a query, inspects the retrieved evidence,
and either continues searching or produces an answer.
The BM25 variant retrieves documents using lexical relevance
scores, whereas the dense variant uses similarity between
query and document embeddings.

\paragraph{DCI-Agent-Lite.}
DCI-Agent-Lite gives the agent direct access to the raw corpus
through general-purpose terminal tools, rather than a separate
retrieval API~\citep{li2026beyond}.
The agent composes lexical searches, locates relevant passages,
and inspects local context to discover clues and verify
candidate answers.

\paragraph{DR-DCI.}
DR-DCI combines retrieval-based candidate discovery with direct
corpus interaction~\citep{lu2026dr}.
Through its \texttt{pull} action, the agent retrieves documents
into a persistent local workspace and then searches and reads
within that workspace.
Additional pulls expand the workspace as new clues or unresolved
evidence needs emerge, allowing retrieval and local investigation
to alternate throughout the search.

\paragraph{RARG.}
RARG uses retrieval relevance to guide grep-style corpus
exploration~\citep{li2026new}.
It orders documents by relevance and directs \texttt{ripgrep}
to traverse them in that order, so that promising documents
are searched before less relevant ones.

\paragraph{PI-SERINI.}
PI-SERINI separates retrieval, result browsing, and document
reading into distinct actions~\citep{hsu2026rethinking}.
It retains a cached ranking that the agent can browse beyond
the initially displayed results without issuing another query.
The agent selectively opens promising documents and reads
relevant portions, allowing deeper retrieval without immediately
placing all retrieved document contents in its context.

\paragraph{\sieve.}
\sieve follows a Search--Inspect--Fetch workflow that preserves
webpage structure during retrieval and evidence
acquisition~\citep{wang2026search}.
The agent uses fielded Boolean queries to select eligible
webpages, which are then ordered by a ranker.
Structured result cards expose headings and other navigation
cues, allowing the agent to inspect the available structure
and fetch selected sections rather than entire webpages.

\newpage
\section{Context and Evidence Measurement}
\label{app:evaluation_metrics}

We distinguish candidate membership, source-text observation, and dynamic context retained by the model. Table~\ref{tab:expanded_accounting} summarizes the definitions and accounting scopes of these quantities.

\subsection{Scoring and accounting}
\label{app:scoring_accounting}

\paragraph{Answer quality.}
For all benchmarks, an LLM judge determines whether the final answer extracted from the agent's response is semantically equivalent to the reference answer. For BC+, we use the grading prompt from the official BC+ evaluation code~\citep{chen2025browsecomp}, reproduced verbatim below. The judge is \texttt{gpt-5.4-nano} with reasoning effort set to \texttt{low}. Accuracy is the percentage of questions receiving a \texttt{correct: yes} verdict.

\begin{iaCBox}{BC+ judge prompt}
\setlength{\parskip}{4pt}
Judge whether the following [response] to [question] is correct or not based on the precise and unambiguous [correct\_answer] below.

[question]: \{question\}

[response]: \{response\}

[correct\_answer]: \{correct\_answer\}

Your judgement must be in the format and criteria specified below:

extracted\_final\_answer: The final exact answer extracted from the [response].

[correct\_answer]: Repeat the [correct\_answer] given above.

reasoning: Explain why the extracted\_final\_answer is correct or incorrect based on [correct\_answer], in the context of this [question]. You should judge whether the extracted\_final\_answer is semantically equivalent to [correct\_answer], allowing the extracted\_final\_answer to be string variations of [correct\_answer]. You should also allow the extracted\_final\_answer to be more precise or verbose than [correct\_answer], as long as its additional details are correct. Do not comment on any background to the problem, do not attempt to solve the problem, do not argue for any answer different than [correct\_answer], focus only on whether the answers are semantically equivalent.

correct: Answer \textquotesingle yes\textquotesingle{} if extracted\_final\_answer matches the [correct\_answer] given above, or is within a small margin of error for numerical problems. Answer \textquotesingle no\textquotesingle{} otherwise, i.e.\ if there if there is any inconsistency, ambiguity, non-equivalency, or if the extracted answer is incorrect.

confidence: The extracted confidence score between 0\textbar\textbackslash\%\textbar{} and 100\textbar\textbackslash\%\textbar{} from [response]. Put 100 if there is no confidence score available.
\end{iaCBox}

For the four multi-hop QA benchmarks, reference answers are concise entities, dates, or numbers, so an LLM judge can determine answer equivalence with little ambiguity while accommodating differences in surface form. Following prior work on agentic search~\citep{li2026beyond}, we use an LLM judge with the grading prompt from the official BrowseComp evaluation code~\citep{wei2025browsecomp}, which also keeps answer evaluation consistent with the LLM-judge protocol used for BC+. The prompt is reproduced verbatim below.

\begin{iaCBox}{Multi-hop QA judge prompt}
\setlength{\parskip}{4pt}
Judge whether the following [response] to [question] is correct or not based on the precise and unambiguous [correct\_answer] below.

[question]: \{question\}

[response]: \{response\}

Your judgement must be in the format and criteria specified below:

extracted\_final\_answer: The final exact answer extracted from the [response]. Put the extracted answer as \textquotesingle None\textquotesingle{} if there is no exact, final answer to extract from the response.

[correct\_answer]: \{correct\_answer\}

reasoning: Explain why the extracted\_final\_answer is correct or incorrect based on [correct\_answer], focusing only on if there are meaningful differences between [correct\_answer] and the extracted\_final\_answer. Do not comment on any background to the problem, do not attempt to solve the problem, do not argue for any answer different than [correct\_answer], focus only on whether the answers match.

correct: Answer \textquotesingle yes\textquotesingle{} if extracted\_final\_answer matches the [correct\_answer] given above, or is within a small margin of error for numerical problems. Answer \textquotesingle no\textquotesingle{} otherwise, i.e.\ if there if there is any inconsistency, ambiguity, non-equivalency, or if the extracted answer is incorrect.

confidence: The extracted confidence score between 0\textbar\textbackslash\%\textbar{} and 100\textbar\textbackslash\%\textbar{} from [response]. Put 100 if there is no confidence score available.
\end{iaCBox}

\paragraph{Average live context.}
Let $H^{\mathrm{dyn}}_{q,t}$ be the dynamic-history text supplied to decision-making model call $t$ for question $q$, after truncation or compaction. We exclude fixed prompts, tool schemas, and the original question. With $T_q$ such calls, the reported quantity is
\begin{equation}
 \operatorname{AvgCtx}=\frac{1}{|Q|}\sum_{q\in Q}
 \frac{1}{T_q}\sum_{t=1}^{T_q}
 \operatorname{tok}(H^{\mathrm{dyn}}_{q,t}).
 \label{eq:expanded_live}
\end{equation}
Token counts use the shared  \texttt{o200k\_base}~\citep{openai2022tiktoken} tokenizer, applied to the recorded readable dynamic-history text. Reasoning tokens are excluded, while any summaries or state information shown to the model are included, even when placed in the system prompt. We average first within each question and then across questions, giving each question equal weight.

\paragraph{Evidence coverage.}
Let $E_q$ be the annotated evidence-document set and $O_q$ the documents whose source text enters at least one agent input. Following the definition in the main text,
\begin{equation}
 \operatorname{EviCov}=\frac{100}{|Q|}\sum_{q\in Q}
 \frac{|E_q\cap O_q|}{|E_q|}.
 \label{eq:expanded_coverage}
\end{equation}
Evidence coverage measures document-level source observation, rather than passage-level evidence use. We count at the document level because BC+ annotates evidence by document and because methods return text in different units, such as passages, snippets, and terminal output; whether observed evidence is used correctly is reflected in answer accuracy. Candidate membership or a filename alone does not count as observation. We use the full BC+ evidence annotations rather than only the gold documents containing the answer, since the metric is intended to capture how much supporting evidence the agent discovers during exploration.

\paragraph{Model calls and source tokens.}
\emph{Calls} is the mean number of decision-making model requests per question. \emph{SrcTok} is the per-question mean of cumulative source-text tokens returned by tool executions. A new tool execution returning previously seen text is counted again, whereas replaying an existing tool result in subsequent model inputs does not add to SrcTok. The preliminary analysis separately measures deduplicated source exposure.

\begin{table}[htbp]
\centering
\vspace{-0.5cm}
\caption{Definitions and accounting scopes of context and evidence measures.}
\label{tab:expanded_accounting}
\begin{iaCTable}
\begin{tabularx}{\linewidth}{@{}l X X@{}}
\toprule
Quantity & What is counted & Accounting scope \\
\midrule
Candidate membership
& Documents in a retained state.
& Membership in the exploration state. \\
Evidence coverage (EviCov)
& Annotated evidence documents with observed source text.
& Document-level source observation. \\
Avg. live context (AvgCtx)
& Readable dynamic-history tokens supplied at each decision.
& Averaged within questions, then across questions. \\
Source delivery (SrcTok)
& Source-text tokens returned by tool executions.
& Cumulative delivery per question, including repeated reads. \\
Unique exposure (Figure~\ref{fig:preliminary_evidence_exposure})
& Observed source intervals merged by document and position.
& Deduplicated exposure within each trajectory. \\
\bottomrule
\end{tabularx}
\vspace{-0.5cm}
\end{iaCTable}
\end{table}

\subsection{Preliminary analysis}
\label{app:preliminary_analysis_details}

We analyze Retrieval Agent (BM25), based on the official BC+ implementation, and DCI-Agent-Lite on the same fixed 100-question subset. Both use GPT-5.4 nano with high reasoning effort. The two analyses use the same runs.

\paragraph{Source provenance.}
We group source text into gold documents, other annotated evidence documents, and documents outside the annotated evidence set. \emph{Other} includes agent-generated messages, non-source tool output, and text whose source attribution cannot be verified.

Returned passages and local search/read outputs are associated with documents using explicit document identifiers for structured tool results and resolved corpus paths for terminal outputs. We then verify the returned text against the corresponding canonical document body. For terminal outputs, verification requires whitespace-normalized body fragments of at least 24 Unicode codepoints. Filenames, metadata headers, and title-only matches do not establish source-text exposure. Token accounting follows Appendix~\ref{app:scoring_accounting}.

\paragraph{Analysis I: retained context.}
We examine the final logged model request after truncation or compaction, excluding the question, fixed instructions, and tool specifications. Category shares are calculated within each request and then averaged across trajectories, giving each trajectory equal weight. This analysis characterizes the composition of the dynamic context retained at the final model request.

\paragraph{Analysis II: first evidence exposure.}
Among trajectories that reach an annotated evidence source, we identify the first model request containing its text. Deduplicated non-evidence source text is divided into content first observed up to and including that request and content first observed afterward. We normalize both amounts by the trajectory's total deduplicated non-evidence text and average the shares for each agent. Non-evidence text in the first evidence-bearing request belongs to the first group.

Repeated content is deduplicated using within-trajectory unions of observed intervals in each canonical document body. Intervals are located using recorded offsets, verified line numbers, or unique whitespace-normalized substring matches. Overlapping intervals are counted only once and assigned to their earliest observed request; newly exposed regions of a previously read document still count. Identical text at different source positions or in different documents is not merged.

These analyses characterize context retention and exposure timing relative to the annotated evidence set.

\newpage
\section{Controlled Variants}
\label{app:additional_analyses}

The ablations examine refinement feedback and candidate-state capabilities separately. Within each comparison, the question set, corpus, backbone, and remaining interface components are held fixed. 
For variants that remove a capability, descriptions of the removed operations are also deleted from the agent instruction (Appendix~\ref{app:prompts}), so that the agent is not directed toward unavailable operations; the remaining instruction text is unchanged.
Both ablation groups use the same fixed 100-question BC+ subset as the preliminary analysis (Appendix~\ref{app:preliminary_analysis_details}). Results are reported in Tables~\ref{tab:observation_ablation} and~\ref{tab:state_ablation}.

\subsection{Refinement feedback and selective reading}
\label{app:observation_analysis}

\paragraph{Auto Preview.}
This variant adds bounded source-text snippets to successfully completed \texttt{FILTER}, \texttt{INTERSECT}, \texttt{UNION}, and \texttt{DIFFERENCE} responses, while preserving candidate-state operations and explicit reading. Each preview selects up to five documents in the resulting set's canonical document-key order, the same order in which an explicit \texttt{READ} of an unordered state returns documents (Appendix~\ref{app:reading_execution}). Previews therefore add no ranking signal: all selected documents belong to the operation's result, and ranking remains available only when the agent explicitly applies \texttt{RANK}. Selecting documents by relevance would instead confound automatic source-text delivery with ranking. Each selected document contributes one snippet of at most 440 Unicode codepoints, yielding at most 2,200 source codepoints per qualifying operation.

Snippet anchors are the first 32 distinct positive lexical terms collected from the operation and its parent-state lineage. Negated terms are excluded, and \texttt{DIFFERENCE} inherits terms only from its left operand. When no lineage terms are available, the first 32 distinct analyzed question tokens are used. The snippet extractor selects a term-matched window, falling back to the document prefix when no terms match. Empty states return no source text.

Previews are attached to the existing tool response without an additional model call. The preview budget applies independently to each qualifying operation within a batch, and repeated or overlapping previews are retained. State references and statistical feedback remain available.

\paragraph{Binary Feedback.}
This variant replaces every explicit count shown to the agent with an empty/non-empty indicator. This applies to state-producing outputs, \texttt{COUNT} and \texttt{COUNT\_DOCS} responses, state information shown to the model (including the corpus state), and read responses, and numerical counts are also removed from error messages.

All operations remain enabled, and source text, document identifiers, offsets, pagination, and read-budget information are unchanged. The variant removes only explicit numerical feedback: the agent could still infer set sizes by paging through documents or issuing repeated binary queries.

\subsection{Candidate-state capabilities}
\label{app:state_analysis}

The state ablations remove one capability at a time while retaining the observation policy of \proposed{}. State operations return references and statistical feedback, and source text is obtained through explicit reading.

\paragraph{Without state reuse.}
This variant disables cross-call reuse of intermediate candidate states for search and refinement. Sequential refinement remains available within a single \texttt{index\_execute} call through batch-local bindings. Search or refinement in a subsequent call must reconstruct the candidate state from \texttt{CORPUS}.

Previously created states remain available for explicit reading, including repeated and paginated reads, plain \texttt{COUNT}, and state inspection. The ablation therefore isolates cross-call search reuse while preserving access to previously selected documents.

\paragraph{Without set composition.}
This variant removes \texttt{INTERSECT}, \texttt{UNION}, \texttt{DIFFERENCE}, and \texttt{RESTRICT}, while preserving Boolean filtering. Removing \texttt{RESTRICT} also closes the ranked-set route for combining retained memberships.

The agent can express multiple conditions within a new filter. For Boolean conditions over the same corpus,
\[
F(\mathcal C,p)\cap F(\mathcal C,q)
=
F(\mathcal C,p\wedge q).
\]
The intervention isolates direct composition of previously computed candidate states while retaining conjunctive filtering.

\paragraph{Without positional operations.}
This variant removes \texttt{PHRASE} and \texttt{NEAR}, while retaining term matching and Boolean conditions. The restriction applies recursively to candidate conditions in \texttt{FILTER}/\texttt{COUNT\_DOCS} and to reading anchors in \texttt{AROUND}. Term-anchored reading and existing term-based snippet extraction remain available, and the underlying positional index is unchanged. The ablation evaluates phrase and proximity constraints across both candidate selection and targeted reading.

\newpage
\section{Corpus Scale and Resource Use}
\label{app:corpus_scaling}

The scaling study keeps 100 BC+ questions and their original evidence fixed while expanding the released corpus of approximately 100K documents to approximately 200K, 400K, and 800K documents with nested samples of FineWeb pages. Each larger collection contains the original corpus and the distractors included at the smaller scale. At a given scale, \proposed{}, Retrieval Agent (BM25), and DCI-Agent-Lite use the same collection. Each method's model settings and execution budgets remain fixed across scales.

\paragraph{API cost.}
Live-context size is not a billing measure: earlier history can be processed at multiple model calls, and cached input can be priced differently from uncached input. API cost is calculated from recorded usage using the experiment's fixed gpt-5.4-nano rates of \$0.20 per million uncached input tokens, \$0.02 per million cached input tokens, and \$1.25 per million output tokens. Cached input is charged at the discounted rate rather than treated as free.

\paragraph{Completion and interpretation.}
Cost should be interpreted alongside task completion, as unfinished runs may consume fewer resources without solving the question.

\section{Agent Instructions}
\label{app:prompts}

The following block reproduces the agent instruction used in all \proposed{} runs.
Formatting is adjusted for readability; the wording is unchanged. The question is supplied separately. Numeric values in the example tool calls illustrate usage rather than define run-wide limits.
The instruction permits reading before the candidate set becomes small; it imposes no fixed count threshold for reading. The answer-judge instruction is specified separately in Appendix~\ref{app:evaluation_metrics}.

\begin{iaCPrompt}
\iaCPromptHead{Objective}
Answer the user's question using evidence from the current immutable corpus snapshot. Treat corpus evidence observed through \texttt{index\_read} as the authoritative basis for factual claims. Do not claim to have seen corpus content that has not appeared in an \texttt{index\_read} result, and never invent an evidence ID.

Finish by calling \texttt{submit\_answer}.

\iaCPromptHead{Mental model}
You work with persistent retrieval states, not transient search-result pages.

\begin{itemize}
\item \texttt{CORPUS} is the root \texttt{SetRef} containing every document in the snapshot.
\item A \texttt{SetRef} is an unordered, persistent document membership set.
\item A \texttt{RankedRef} has persistent membership plus a deterministic reading order.
\item State-producing operations create a new immutable state and return its exact cardinality.
\item Existing states remain available across turns. Reuse them, branch from them, and recombine them.
\item State handles such as \texttt{s3} and \texttt{r2} persist across tool calls.
\item A batch \texttt{binding} exists only inside one \texttt{index\_execute} call. Use the persistent handle returned in the result on later turns.
\item Search, counting, ranking, and state-control operations do not return corpus text.
\item Only \texttt{index\_read} can place corpus text in your context.
\end{itemize}
The runtime may show a deterministic state ledger and evidence ledger. They are memory aids containing only previously observed state metadata and evidence. They are not new corpus evidence.

\iaCPromptHead{Expressions and states}
\iaCPromptHead{Lexical expressions}
\begin{itemize}
\item \texttt{TERM("x")} matches one analyzed token. It is not substring, wildcard, or regex search.
\item \texttt{PHRASE(...)} matches ordered, adjacent lexical expressions.
\item \texttt{NEAR(...)} matches lexical expressions under the specified order and maximum-gap constraint.
\item \texttt{ANY\_OF(...)} is positional choice between lexical expressions and may be used where match locations matter.
\end{itemize}
\iaCPromptHead{Boolean conditions}
\begin{itemize}
\item \texttt{AND}, Boolean \texttt{OR}, and \texttt{NOT} combine unexecuted document conditions inside \texttt{FILTER} or \texttt{COUNT\_DOCS}.
\item Boolean \texttt{OR} is document-level truth. It is different from positional \texttt{ANY\_OF}.
\end{itemize}
\iaCPromptHead{Persistent-state algebra}
\begin{itemize}
\item \texttt{INTERSECT}, \texttt{UNION}, and \texttt{DIFFERENCE} combine already computed \texttt{SetRef} states.
\item Use condition algebra when composing one predicate before execution.
\item Use state algebra when independently computed intermediate results should remain reusable or branchable.
\end{itemize}
\iaCPromptHead{Eligibility and ordering}
Keep hard eligibility separate from soft reading order.

\begin{itemize}
\item \texttt{FILTER} changes membership by applying an exact lexical or Boolean condition to a \texttt{SetRef}.
\item \texttt{COUNT\_DOCS} tests a condition within an existing state without creating another state.
\item Most state-producing results already include cardinality; use \texttt{COUNT} when you need to re-read an existing state's cardinality.
\item \texttt{RANK} preserves every input member and adds an order from soft lexical evidence.
\item \texttt{TOPK} keeps a prefix of a \texttt{RankedRef}.
\item \texttt{RESTRICT} keeps only documents allowed by a \texttt{SetRef} while preserving an existing ranked order.
\item \texttt{AS\_SET} deliberately discards rank and preserves membership.
\end{itemize}
Do not use a soft clue as a hard filter merely because it may be relevant. Conversely, do not rank on a condition that must be true for the answer.

\iaCPromptHead{Native tools}
\iaCPromptHead{\texttt{index\_execute}}
Use \texttt{index\_execute} for ordered batches of non-text operations:

\begin{lstlisting}[style=iaCCode]
FILTER, INTERSECT, UNION, DIFFERENCE,
COUNT, COUNT_DOCS,
RANK, TOPK, RESTRICT, AS_SET
\end{lstlisting}
Use one batch when later operations depend on earlier results. Batch-local bindings avoid extra model turns, but they do not replace the persistent handles returned by state-producing results.

Example: narrow a state and inspect its size without reading text.

\begin{lstlisting}[style=iaCCode]
{
  "operations": [
    {"op": "FILTER", "target": "CORPUS", "condition": {"term": "mars"},
     "bind": "mars_docs"},
    {"op": "FILTER", "target": {"binding": "mars_docs"},
     "condition": {"term": "crater"}, "bind": "candidates"},
    {"op": "COUNT", "state": {"binding": "candidates"}}
  ]
}
\end{lstlisting}
Example: preserve alternative hypotheses and recombine them.

\begin{lstlisting}[style=iaCCode]
{
  "operations": [
    {"op": "FILTER", "target": {"handle": "s1"},
     "condition": {"term": "volcano"}, "bind": "volcanic"},
    {"op": "FILTER", "target": {"handle": "s1"},
     "condition": {"term": "impact"}, "bind": "impact"},
    {"op": "UNION", "left": {"binding": "volcanic"},
     "right": {"binding": "impact"}, "bind": "combined"}
  ]
}
\end{lstlisting}
\iaCPromptHead{\texttt{index\_read}}
\texttt{index\_read} is the only tool that materializes corpus text.

Use it deliberately when text is likely to reveal:

\begin{itemize}
\item a fact needed for the answer;
\item a bridge entity for another search hop;
\item the corpus's actual vocabulary;
\item evidence needed to distinguish or verify hypotheses.
\end{itemize}
You do not need to wait for a fixed cardinality before reading. Balance the expected information value of text against its context cost.

Reading modes:

\begin{itemize}
\item \texttt{DOCUMENT}: read a complete selected document.
\item \texttt{AROUND}: read bounded windows around lexical occurrences.
\item \texttt{RANGE}: read an exact raw-text range and is valid only for a direct \texttt{doc\_key} target.
\item For a state target, use a bounded \texttt{PAGE} unless you deliberately intend to process every document.
\end{itemize}
Example: read two ranked or set-state documents around the first occurrence of a lexical anchor.

\begin{lstlisting}[style=iaCCode]
{
  "target": {"state": {"handle": "s7"}},
  "documents": {"page": {"limit": 2, "after": null}},
  "region": {"around": {
    "anchor": {"term": "crater"}, "selector": {"first": {}},
    "before": 120, "after": 120
  }},
  "budget": {"max_output_codepoints": 4000, "max_evidence_count": 8}
}
\end{lstlisting}
A \texttt{ReadBudgetExceeded} result contains no partial evidence text. Use its structured diagnostics and page-selection metadata to choose among:

\begin{itemize}
\item retrying with a smaller page or region;
\item reading a selected \texttt{doc\_key} with a bounded \texttt{RANGE};
\item advancing past the selected page when appropriate.
\end{itemize}
\begin{samepage}
\iaCPromptHead{\texttt{index\_state}}
Use \texttt{index\_state} to:

\begin{itemize}
\item list active states when the runtime ledger is insufficient;
\item inspect a lineage node when derivation provenance matters;
\item release states that are no longer useful.
\end{itemize}
Do not release a state that may still be needed for a branch or later recombination.

\end{samepage}
\iaCPromptHead{\texttt{submit\_answer}}
Call \texttt{submit\_answer} only when the answer is sufficiently supported by observed corpus evidence or when you have exhausted the allowed search and must report the best supported result.

\begin{itemize}
\item Include the IDs of Evidence that directly support the answer.
\item Cite only Evidence IDs actually returned by prior \texttt{index\_read} calls.
\item Do not cite state handles as evidence.
\item Do not invent document text, ranges, or evidence IDs.
\end{itemize}
\iaCPromptHead{Search policy}
Use cardinality and observed evidence as feedback, not as rigid rules.

\begin{itemize}
\item A zero-result state can mean that a surface form is absent, the analyzer produced a different vocabulary item, or the condition is too specific. Revise the relevant hypothesis rather than blindly repeating it.
\item A very broad state may benefit from another hard condition, an explicit branch, set recombination, or soft ranking.
\item A manageable state may be read directly, but there is no universal size threshold.
\item Read earlier when new vocabulary or a bridge entity is needed to formulate the next predicate.
\item After reading, use newly observed vocabulary to refine the most relevant existing state or create a deliberate new branch.
\item Preserve alternative hypotheses as separate states when later evidence may distinguish them.
\item Reuse an existing state instead of rerunning an equivalent derivation.
\item Inspect structured tool errors and correct the specific invalid reference, type, value, cursor, or budget. Do not repeat the same invalid call unchanged.
\end{itemize}
The objective is not to minimize READ calls at all costs. The objective is to control when evidence enters context while using persistent index state to test and revise search hypotheses efficiently.

\end{iaCPrompt}

\clearpage
\endgroup

\end{document}